\documentclass[useAMS,amssymb,epsfig,usegraphicx,usenatbib,twocolumn,revtex4-1]{emulateapj}

\def\aj{AJ}% Astronomical Journal

\def\apj{ApJ}% Astrophysical Journal
\def\apjl{ApJ}% Astrophysical Journal, Letters
\def\aap{A\&A}% Astronomy and Astrophysics
\def\mnras{MNRAS}% Monthly Notices of the RAS
\def\nat{Nature}% Nature\def\pasj{PASJ} % Publications of Astronomical Society Japan
\def\na{NewA}% New Astronomy
\def\pasp{PASP} % Publications of Astronomical Society of the Pacific
\def\pasa{PASA} % Publications of Astronomical Society of Australia
\usepackage[usenames,dvips]{color}

\newif\ifAMStwofonts

\makeatother

\shorttitle{}
\shortauthors{Saha \& Cortesi}

\begin{document}

\title{Forming Lenticular Galaxies via Violent Disk Instability}
\author{Kanak Saha$^{1}$ \& Arianna Cortesi$^{2}$}
\affil{$^{1}$ Inter-University Centre for Astronomy and Astrophysics, Pune 411007, India, \\
$^{2}$ IAG - Universidade de São Paulo, Brazil
\\e-mail:kanak@iucaa.in, aricorte@googlemail.com}

\label{firstpage}

%\maketitle

\begin{abstract}
Lenticular galaxies are generally thought to have descended from spirals via morphological transformation, 
although recent numerical simulations have shown that minor or even major merger can also lead to an S0-like remnant. These mechanisms, however, are active in a dense environment such as a group or a cluster of 
galaxies - making it harder to explain the remarkable fraction of S0s found in the field. 
Here, we propose a new mechanism 
to form such lenticular galaxies. We show that an isolated cold disk settled into rotational equilibrium becomes 
violently unstable - leading to fragmentation and formation of stellar clumps that, in turn, not only grow the bulge, 
but also increase the stellar disk velocity dispersion optimally in less than a billion year. Subsequently, the galaxy evolves passively without any conspicuous spiral structure. The final galaxy models resemble  
remarkably well the morphology and stellar kinematics of the present-day S0s observed by the Planetary Nebulae spectrograph. Our findings suggest a natural link between the high-redshift clumpy progenitors to the 
present-day S0 galaxies.

\end{abstract}


\section{Introduction}
\label{sec:intro}

The broad morphological features of lenticular (S0) galaxies are analogous to spiral
galaxies, except that S0s are devoid of any conspicuous spiral structures. 
Traditionally, lenticular galaxies were thought to be disk galaxies dominated by a
bulge component with bulge-to-total ratio $B/T > 0.6$ \citep{Simien-DeVaucouleurs1986}. However, in
the light of recent studies of bulge-disk decomposition on a large number of disk galaxies,
such a scenario has been diversified  - S0 galaxies are now known to host bulges
whose structural properties are as similar as in spiral galaxies e.g., their Sersic indices are 
found to vary from $n=1$ to $n=4$ or so; the $B/T$ can be as low as $0.1$ - as in late-type spirals \citep{Aguerrietal2005,Laurikainenetal2005,Weinzirletal2009, Laurikainenetal2010,Barwayetal2016}. 
In view of these findings, it is apparent that bulges of S0s share similarity with those in spiral 
galaxies, implying that the formation and growth of bulges might be irrespective of the host
disk properties. 

Not only in terms of morphology, S0s also share kinematic similarity with spirals to a large extent
 e.g., they follow Tully-Fisher relation albeit with some offset \citep{Williamsetal2010, Cortesietal2013b}. 
 Detailed kinematics derived from Planetary Nebulae observations indicate that S0 disks are, in general, 
 hotter, with $V_{\varphi}/\sigma_{\varphi}$ smaller by a factor of few than those of normal spirals, 
 where $V_{\varphi}$ and $\sigma_{\varphi}$ are the mean azimuthal velocity and dispersion of stars respectively\citep{Cortesietal2013b}. Despite having such morpho-kinamtic similarity, it remains unclear whether S0s 
 have formed in the early universe and evolved rather passively or they have undergone morphological transformation to become one of the present-day S0s.
 
Based on the close analogies between S0s and spirals, it has been suggested that S0s
have descended from spirals via morphological transformation. A number of physical
processes are being proposed as the primary drivers of such transformation e.g., gas 
removal during galaxy collisions \citep{Spitzer-Baade1951}; ram pressure 
stripping of gas and dust from spirals \citep{GunnGott1972, Larsonetal1980}; galaxy harassment \citep{Mooreetal1996} or gas starvation leading to star-formation shut down \citep{Bekkietal2002, Pengetal2015}. 
Apart from these mechanisms, numerical simulations have shown that major mergers (1:1 or even 3:1 mass-ratios)
 or dry minor mergers can lead to S0 like remnants or even 
transform spirals to S0s \cite[see][]{Bekkietal2011,Eliche-Moraletal2013, Tapiaetal2014, Querejetaetal2015}. 
Most of these mechanisms, however, are active in dense environment such 
as galaxy clusters or group environment - making it harder to explain $\sim 30\%$ of field lenticulars \citep[see][]{vandenbergh2009}. Comparing the K-band luminosities between S0s and spirals, \cite{Bursteinetal2005} suggested that S0s are not gas-stripped spirals. Besides,\cite{vandenbergh2009} pointed out that the observed distribution of flattening is found to be independent of environment. These evidences indicate that perhaps other mechanisms 
are also in place behind making the present-day S0s, especially ones in the field.

In this letter, we propose violent disk instability \citep{Toomre1964,Ceverinoetal2017} and fragmentation as one of the possible mechanisms to produce S0 galaxies in numerical simulations. We use isolated collisionless simulations to capture the basic underlying physics that governs the formation of S0 galaxies and compare their morphology and kinematics with observed ones.  

\begin{figure}
\rotatebox{0}{\includegraphics[width=0.4\textwidth]{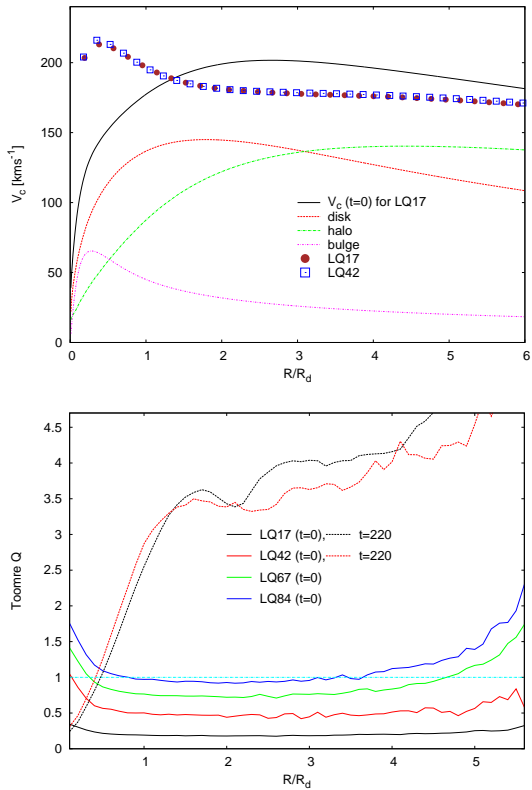}}
\caption{Top: Circular velocity curves for model LQ17 at t=0 and the contribution from disk, 
bulge and halo. Overplotted are the two circular velocity curves at t=220 for models LQ17 and LQ42.
Bottom: Initial Toomre Q profiles for the stellar disk for all the models. Dashed lines 
show the Toomre Q profiles at t=220
for LQ17 and LQ42.}
\label{fig:VcQ}
\end{figure}  

\begin{figure}[b]
\centering
\rotatebox{0}{\includegraphics[width=0.5\textwidth ]{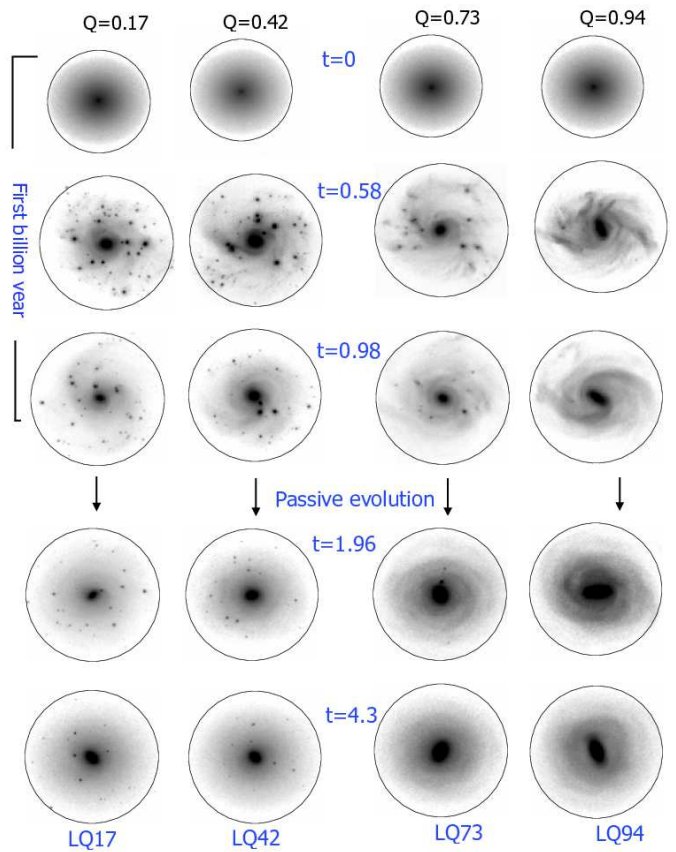}}
\caption{Stellar surface density in near face-on projection  (at inclination angle of 30 deg)
at different epochs for LQ17, LQ42, LQ73 and LQ94. The time between two successive snapshots is $19.6$~Myr. The quoted initial Q value is at $2.5 R_d $. The simulated images were convolved with a PSF that is ideal for S-PLUS.}
\label{fig:surfden}
\end{figure}

\begin{figure*}
\rotatebox{0}{\includegraphics[width=0.8\textwidth]{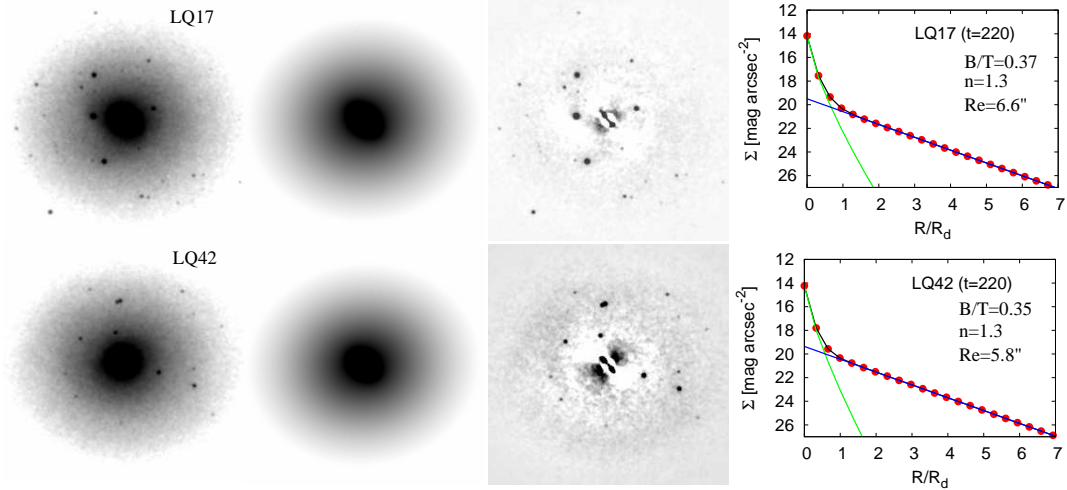}}
\caption{Two-component bulge-disk decomposition of the final galaxy models: {\it first coloumn:} images, {\it second:} model, {\it third:} residual, {\it and fourth:} surface brightness profile. The blue curve shows the disk light, the green one the spheroid profile, the black curve represent the total light and the red dots represent the stellar density. The simuated galaxies are kept at a distance of 15.4 Mpc and 1"=0.074 kpc. }
\label{fig:galfit}
\end{figure*}
  
\section{Model setup and simulation}
\label{sec:model}

We construct a set of $4$, three-component galaxy models consisting of a stellar disk, 
a dark matter halo, and a classical bulge, initially in equilibrium. The stellar disk 
is initially axisymmetric with surface density following an exponential profile. 
The initial vertical density follows a $sech^{2}$~distribution with a constant scale height.
For self-consistency, we let the vertical velocity dispersion follow also an exponential 
profile with a scale length, $R_{\sigma} = 2 R_d$, where $R_d$ is the disk scale length. The dark matter halo 
is modelled with a lowered \citep{Evans1993} model, which generates a nearly flat circular velocity 
profile (Fig.~\ref{fig:VcQ}) and the bulge with a King DF \citep{King1966}. Further details about model construction can be found in \cite{KD1995} and \cite{Sahaetal2010}. The parameters describing the mass models of stellar disk, dark matter halo and the initial bulge are kept identical in all the models. What has been varied is the initial radial velocity dispersion ($\sigma_r$) of the stars in the disk. This, in turn, characterises each model with a particular Toomre Q profile, defined as $Q (R) = \frac{\sigma_r (R) \kappa (R)}{3.36 G \Sigma(R)}$ 
\noindent where $\kappa$ is the epicyclic frequency and $\Sigma$ is the surface density. The
initial $Q$-profiles, for all the four models LQ17, LQ42, LQ73 and LQ94, are shown in Fig.~\ref{fig:VcQ} and they are subject to strong axisymmetric instability \citep{Toomre1964}.
 
The models are scaled such that $R_d = 3$~kpc and the circular velocity at $2 R_d$ is
$200$~km/s. The disk mass, $M_{disk} = 4.6 \times 10^{10} M_{\odot}$, the bulge mass 
$M_{bulge} = 0.13 \times 10^{10} M_{\odot}$ and halo mass, $M_{halo} = 8.4 \times 10^{10} M_{\odot}$. 
The initial bulge-to-total ratio is given by $B/T = 0.03$, same for all the models.
Note that our galaxy models are disk dominated, in sync with recent observation of lenticular galaxy 
NGC 3998 \citep{Boardmanetal2016}. We have used a total of $3.7 \times 10^{6}$ particles. It has been 
shown by a number of studies that convergence in terms of discreteness noise is reached with a few million particles \citep{Dubinskietal2009,Sahaetal2010, SahaElmegreen2018} as we have chosen for our models. The softening lengths for the disk, bulge and halo particles are unequal and are calculated following the suggestion of \cite{McMillan2007}. The simulations are performed using the 
Gadget code \citep{Springeletal2001} with a tolerance parameter $\theta_{tol} =0.7$ and an integration time step of $0.03$ times the internal time unit. Each model was evolved for a time period of $\sim 4.3$~Gyr. 

\section{Cold Disk Evolution - S0-like morphology}
\label{sec:evolution}

The linear stability analysis shows that an initially axisymmetric, rotationally supported stellar 
disk where stars undergo epicyclic motion become violently unstable when the disk is cold, 
with Toomre $Q < 1$ \citep{Toomre1964,BT1987}. Depending on the minimum Q value, the disk may undergo fragmentation, leading to the the formation of stellar clumps. 
Fig.~\ref{fig:surfden} depicts the morphological evolution of all four models, two with $Q< 0.5$ and two with 
$Q> 0.5$. The model LQ17 being coldest, suffers the strongest instability and produce the largest number of 
stellar clumps. While models with relatively higher $Q$, fragments mildly and make fewer and fewer 
clumps as in LQ73. In fact, LQ94, with $Q \simeq1$ does not even undergo any 
fragmentation and evolves into a typical barred galaxy at 
the end of the simulation. The model LQ73 follows LQ94 but with a lowered bar strength. In the rest of the paper, we mainly concentrate on models LQ17 and LQ42 as potential unbarred S0 candidate, 
as the bar fraction in lenticulars are generally low, \citep[see][]{Butaetal2010,Barwayetal2011}.  In both these models, the stellar clumps are subject to dynamical friction \citep{Chandra1943}, where the frictional force is 
$F_{dyn} \propto M_{clump}^2$, $M_{clump}$ being the clump mass. According to this, the massive stellar clumps are subject to strong dynamical friction and in-spiral to the central region within a billion year time-scale to grow the bulge, see Fig.~\ref{fig:surfden} 
at $t=0.58, 0.98$~Gyr snapshots for visual impression. Soon the massive clumps disappear to the central region, leaving a smooth disk with several lower mass stellar clumps which take much 
longer time to migrate to the centre. In fact, for the lower mass clumps that formed in the outskirts 
of the disk, it would take several billion years to migrate to the centre and hence might still be lingering 
in the disk to be observable (provided having the adequate telescope resolution and image depth). Towards the end of the simulations, both LQ17 and LQ42 evolve in a self-similar fashion achieving a similar morphology - in that 
they both have a smooth disk without any spiral structure, with an enhanced bulge and a number of lower 
mass stellar clumps. They resemble well the observed morphology of S0 galaxies in the field e.g., NGC 7457. 
Connecting this scenario in the high redshift clumpy galaxies crucially depends 
on the clump survival. In our collisionless simulations with no gas dissipation and 
feedback, the clumps are not destroyed - in accordance with recent cosmological 
simulations with supernovae and radiation feedback \citep{Ceverinoetal2017} as well as previous ones 
\citep[e.g.,][]{BournaudElm2007}. But see \cite{Oklopcicetal2017} for a different view.  

In the following sections, we discuss their bulge-disk properties, disk kinematics and compare them with the S0 sample presented in \cite{Cortesietal2013b}.

\begin{figure*}
\begin{center}
\rotatebox{0}{\includegraphics[width=0.7\textwidth]{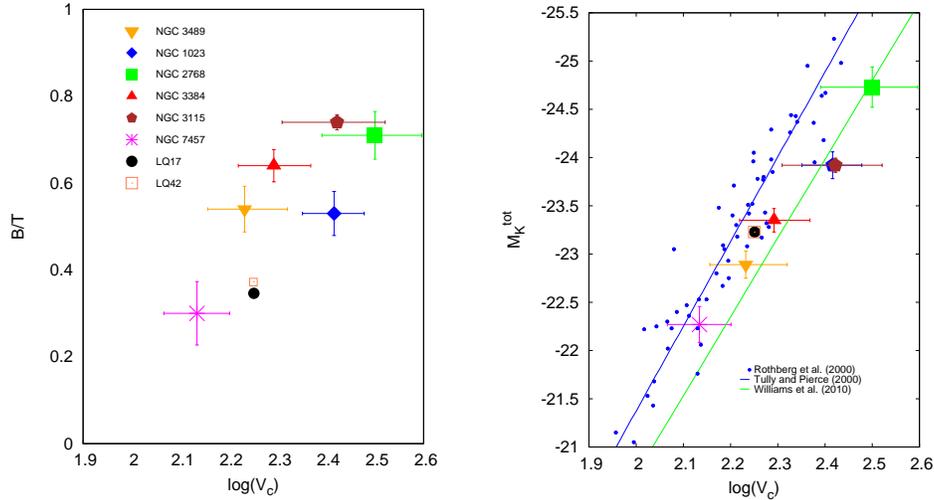}}
\caption{Comparison of our simulated galaxy models with observed S0s \citep{Cortesietal2013b}. {\it Left panel}: variation of $B/T$ ratio with the circular velocity. {\it Right panel}: Tully Fisher relation. blue dots represent spiral galaxies. Blue and green lines are fit to the spirals and massive S0s respectively. The simulated galaxies have an absolute magnitude of $-23.2$ in the K-band.}
\label{fig:kinemat1}
\end{center}
\end{figure*}

\section{bulge-disk decomposition and bulge growth}
\label{sec:decomposition}
We first convolved the simulated images with a PSF that is typical for S-PLUS (Mendes de Oliveira et al. in prep)  observations, specially of an R-band image of the Hydra cluster. Then, we added Poisson noise and a flat sky to these convolved images. We use GALFIT \cite{Pengetal2002} on these simulated images to decompose them 
into two components - a bulge and a stellar disk. The bulge is modelled with a Sersic profile:

\begin{equation}
I_{b}(R) = I_e \, \mathrm{exp}\left\{-b_n[(R/R_e)^{1/n} -1]\right\},
\end{equation}

where $R_e$ is the effective radius, $I_e$ is the effective surface brightness, $n$ is the sersic index and  $b_n = 1.9992n-0.3271$ \cite{GD05}. The stellar disk follows an exponential profile:
\begin{equation}
I_{d}(R) = I_0 \, \mathrm{exp}\left(-\frac{R}{R_d}\right),
\end{equation}
where $I_0$ is the central surface brightness, $R_d$ is the scale length.

The fit is performed for all the time steps of the simulation as presented in Fig.~\ref{fig:surfden}. The first snapshot (t=0) is well fitted by an exponential disk only, i.e. the recovered bulge is three magnitude fainter than the disk, as expected given how the simulation was set. 
In Fig.~\ref{fig:galfit}, we show the decomposition of LQ17 and LQ42. In both cases, the models fit well the smooth component, and the clumps are visible in the residual image, in higher number for the LQ17. The surface brightness profiles resemble remarkably well the profiles recovered for lenticular galaxies with relatively low B/T \citep{Cortesietal2013b}. {\it The clumps that migrated to the centre due to dynamical
friction have grown the initially negligible bulge with $B/T \sim 0.03$, to a final one with 
$B/T \sim 0.37$ and $n=1.3$.} The final effective radii of the bulge is about $0.5$~kpc while the disk scale length is $2.3$~kpc. The sersic indices of the bulges indicate a pseudobulge-like morphology formatioj of which is similar to the way young stellar clumps are believed to have grown bulges in high redshift disks \citep{Noguchi1998,Elmegreenetal2008,InoueSaitoh2012,Bournaud2016}. Interestingly, much of the bulge growth happened during the early phase of evolution and afterwards they remained nearly constant with time, for both simulations. What our mechanism is yet to establish is whether it can create an S0 with a even bigger bulge - more like the classical ones - but this remains the subject of future investigation.

When plotted on the  $B/T$ - $V_c$ plane (see the left panel of Fig.~\ref{fig:kinemat1}), our models seem to be consistent with field lenticulars such as NGC 7457. The nearly flat color gradient and old stellar population observed in many 
field and low-density-environment S0s \citep{Cortesietal2011, Tabor2017}, probably indicate that both the bulge and 
the disk share the same stellar population. In contrast, the bulges of cluster S0s may contain younger stellar population than their disks \citep{Johnston2014}. Since the bulges in our simulated S0s are made from the disk material, they 
would resemble field S0s more than cluster S0s in terms of stellar population and color.

\begin{figure}[b]
\begin{flushleft}
\rotatebox{0}{\includegraphics[width=0.5\textwidth ]{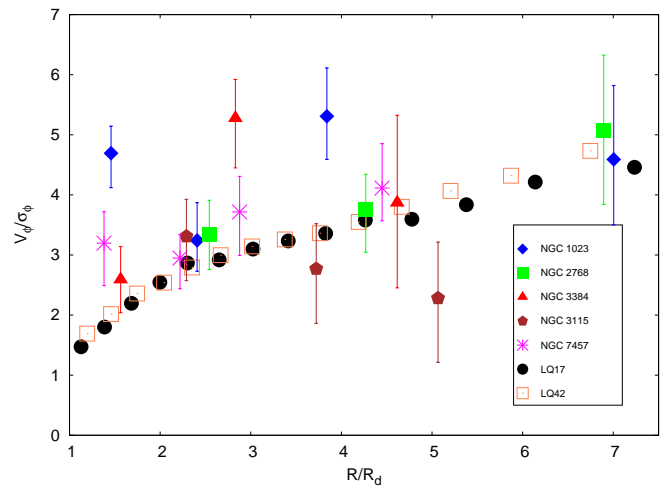}}
\caption{Intrinsic stellar kinematics: simulated radial profiles of $V_{\varphi}/\sigma_{\varphi}$ are compared 
with observed S0s \citep{Cortesietal2013b}. Note NGC~2768 is a group S0 while NGC~7457 from field - display little
difference in stellar kinematics.}
\label{fig:kinemat2}
\end{flushleft}
\end{figure}

\section{Disk stellar kinematics}
\label{sec:kinematics}
The right panel of Fig.~\ref{fig:kinemat1} shows the Tully-Fisher relation of our simulated
galaxies and the observed ones. We notice that LQ17 and LQ42 closely follow the comparatively low-mass S0s (e.g., NGC 7457) on the Tully-Fisher plane. Interestingly, NGC 7457 has a unimodal Globular Cluster population \citep{Hargis2011}, which shares the stellar disk kinematics (Zanatta et al. submitted to MNRAS). In spiral galaxies, red GCs are believed to share the spheroid kinematics, while in elliptical galaxies they generally follow the  kinematics of the overall stellar population. One explanation for the observed disk-like behaviour of  red (or all, for the case of NGC 7457) GCs in some lenticular galaxies, is that they have formed at $z\simeq2$ in star forming clumps, during the gas rich phase of galaxy evolution \citep{Shapiroetal2010, Cortesietal2016}, consistent with the formation mechanism proposed in this work. However, these field low-mass S0s seem to deviate from the TFR followed by the massive S0s \citep{Williamsetal2010}. Considering the TFR as one of the local benchmark scaling relation, massive lenticulars might have a different formation scenario than the low-mass ones. A similar conclusion has also been drawn from the studies of disk-bulge photometric scaling relation \citep[see][]{Barwayetal2007}.

In Fig.~\ref{fig:kinemat2}, we show the radial profile of $V_{\varphi}/\sigma_{\varphi}$ for LQ17 and LQ42 at $t=4.3~Gyr$. In both models, over a wide range of radii, 
$V_{\varphi}/\sigma_{\varphi} \simeq 3 - 4$ - that matches well with a number
of S0 galaxies such as NGC 2768, NGC 3384, NGC 1023, NGC 7457 for which intrinsic 
stellar kinematics \citep{Cortesietal2011,Cortesietal2013b} were derived from observations of 
planetary nebulae using Planetary Nebula Spectrograph \citep{Douglasetal2002, 
Douglasetal2007}. It is interesting to note that the rotation velocities in these 
two models have not reduced significantly with time but the random motion has increased by several
times over its initial value - resulting in a reduced $V_{\varphi}/\sigma_{\varphi}$ value. 
Such a high rise in the stellar velocity dispersion has been possible through the gravitational 
scattering of  the stellar clumps with the background stars - much like the way star-cloud scattering 
increases stellar velocity dispersion in the galactic disk \citep{Spitzer-Schwarzchild1953,Lacey1984}. 
{\it We notice that in model LQ17, the radial velocity dispersion calculated at $R=2 R_d$ increases by 
a factor of $13$ over its initial value within $1$~Gyr and stays nearly constant during the subsequent 
phases of evolution.} Whereas in model LQ42, $\sigma_{r}$ increases moderately by a factor of $\sim 5$ 
within the same time span. In comparison, the heating due to spiral arms alone is rather milder, the radial velocity 
dispersion may increase at the most, by a factor of $2 - 3$ over its initial value on secular evolution 
time scale ($\sim$~few~billion years), 
see \cite{Sahaetal2010}. Note that in order to heat the disk stars, the spiral has to be transient or stochastic 
in nature, as two-armed steadily rotating density waves are unable to heat the disk \citep{BT1987}. 
 This might indicate that spiral arms alone could not have produced such high stellar velocity dispersion that we 
 known from observational modelling of lenticular galaxies \citep{Cortesietal2011}.
\vspace{0.2cm} 
\section{Conclusions}
\label{sec:conclusions}

{\bf 1.} Two of our simulations with live dark halo demonstrate that fragmentation of a low-Q disk ($Q < 0.5$) is a viable alternative to make an S0-like galaxy that are found in the field environment where external trigger or interactions are generally absent. 

{\bf 2.} In both simulations, the model galaxies had no significant spiral structure to begin with. They are able to explain the intrinsic stellar kinematics of field S0s. The other two simulations, having same mass model but with $Q > 0.5$ evolved into typical barred galaxies. 

{\bf 3.} Disk fragmentation and thereby the star-clump scattering are shown to increase the stellar velocity dispersion by a factor of few to $\sim 10$ times its initial value within a Gyr time scale. Within this time period, the stellar clumps that migrate to the central region are shown to have grown the bulge mass by a factor $\sim 10$. 

{\bf 4.} After the first Gyr of rapid evolution, these two models evolved rather passively without any significant 
change, neither in morphology nor in kinematics.

\vspace{-0.2cm}
\section*{acknowledgement} 
All simulations were carried out at the IUCAA~HPC cluster. The authors thank the referee for insightful 
comments.

%===================================================================

\end{document}